\documentclass[aps,11pt,superscriptaddress,onecolumn]{revtex4}
\usepackage{graphicx}
\graphicspath{{./}{./IMAGES/}}
\usepackage{epstopdf}
\usepackage{amssymb}
\usepackage{mathrsfs}
\usepackage{amsmath}
\usepackage{color}
\usepackage{latexsym}
\usepackage{amsfonts}
\usepackage{hyperref}
\usepackage{subfigure}
\usepackage{wasysym}
\usepackage{dcolumn}
\usepackage{bm}
\usepackage[percent]{overpic}
\usepackage{natbib}
\usepackage{booktabs}
\usepackage{gensymb}

\begin{document}

\selectfont

\title{Coarsening Kinetics of Complex Macromolecular Architectures in Bad Solvent}

\author{Mariarita Paciolla}
\affiliation{Centro de F\'isica de Materiales (CSIC, UPV/EHU) and Materials Physics Center MPC, Paseo Manuel de Lardizabal 5, 20018 San Sebasti\'an, Spain}

\author{Daniel J. Arismendi-Arrieta}
\affiliation{Donostia International Physics Center, Paseo Manuel de Lardizabal 4, 20018 San Sebasti\'an, Spain}

\author{Angel J. Moreno}
\affiliation{Centro de F\'isica de Materiales (CSIC, UPV/EHU) and Materials Physics Center MPC, Paseo Manuel de Lardizabal 5, 20018 San Sebasti\'an, Spain}
\affiliation{Donostia International Physics Center, Paseo Manuel de Lardizabal 4, 20018 San Sebasti\'an, Spain}

\begin{abstract}
This study reports a general scenario for the out-of-equilibrium features of collapsing  polymeric architectures. We use molecular dynamics simulations to characterize the coarsening kinetics, in bad solvent, for several macromolecular systems with an increasing degree of structural complexity. In particular, we focus on: flexible and semiflexible polymer chains, star polymers with 3 and 12 arms, and microgels with both ordered and disordered networks. 
Starting from a powerful analogy with critical phenomena, we construct a density field representation that removes fast fluctuations and provides a consistent characterization of the domain growth. Our results indicate that the coarsening kinetics presents a scaling behaviour that is independent of the solvent quality parameter, in analogy to time-temperature superposition principle. Interestingly, the domain growth in time follows a power-law behaviour that is approximately independent of the architecture for all the flexible systems; while it is steeper for the semiflexible chains. Nevertheless, the fractal nature of the dense regions emerging during the collapse exhibits the same scaling behaviour for all the macromolecules. This suggests that the faster growing length scale in the semiflexible chains originates just from a faster mass diffusion along the chain contour, induced by the local stiffness. 
The decay of the dynamic correlations displays scaling behavior with the growing length scale of the system, which is a characteristic signature in coarsening phenomena.
\end{abstract}

\maketitle

\section{Introduction}
Understanding the collapse of, fully polymeric, topologically complex objects in a bad solvent is of broad importance, 
because of its relevance in the early stages of the protein folding \cite{Sadqi2003,Majumder2019}, and its connections with arresting processes or aging phenomena \cite{Majumder2017,Midya2015}.
The thermodynamic and structural properties of the macromolecular collapse in solution, occurring when the quality of the solvent is decreased below a critical value, have been exhaustively investigated over the years \cite{deGennes1979, Kamerlin2016,Grosberg1988,terashima2011,huerta2013,kang2015}, and at equilibrium, the dynamic and static behavior are well known above and below the volume phase transition \cite{Doi1986}. Comparatively, a general framework  for
the non-equilibrium aspects, as the kinetics of the collapse, is still lacking.
The development of new experimental setups for small angle X-ray scattering or single-molecule fluorescence spectroscopy allows to monitor the collapse of a single molecule \cite{Pollack2001,hofmann2012,Zhang2014}, sheding light on the aspects controlling the collapse dynamics.  

Recent computational works have proposed analogies between the collapse of linear polymer chains in bad solvent \cite{Majumder2015EPL,Majumder2017,Majumder-review} and phase-ordering phenomena or coarsening systems (as foams \cite{Saint-Jalmes2006} or polymer blends \cite{Veenstra1999}), where coarsening refers to any out-of-equilibrium relaxation process involving the growth of two separated phases from an initially homogeneous mixture. In a coarsening system the characteristic length scale grows over time and exhibits key universal features like dynamic scaling and self-similarity \cite{Smerlak2018}. In the context of macromolecular collapse in bad solvent, the coarsening process initiates with the formation of small clusters of monomers along the polymeric strands. Afterwards, these clusters become stable and start to grow by withdrawing monomers from the bridges connecting them or by coalescence with other clusters. This moves forward until all the monomers pile up to a single cluster. Finally, the single cluster collapses to the ultimate equilibrium state, characterized by a compact object of melt-like density \cite{Majumder2017}. 
The characteristic length scale of the coarsening process, associated to the collapse from the initial
self-avoiding conformations of the polymer strands to the late fully-collapsed state,
is given by the size of the growing aggregates of monomers.

Some theoretical approaches have been developed for describing the different scaling regimes associated to the coarsening kinetics. 
In binary liquids, where the chemical potential gradient acts as the driving force \cite{Bray2002}, the domains grow with time as $\sim t^{1/3}$.
For fluids and polymers, where hydrodynamic contributions are relevant, linear growth is predicted 
at late times in the viscous regime \cite{Siggia1979} . 
An apparent sublinear regime for the domain growth, $\sim t^{1/2}$, has been
observed at intermediate times in simulations of a gas-liquid separating system \cite{Testard2014}, suggesting an effective interpolation between the former early and late regimes.
In a previous work by some of us \cite{Moreno2018} a scaling exponent of $\sim 0.7$ has been found at intermediate times for the growth of the domains during the deswelling of microgels in bad solvent. This accelaration (higher exponent) with respect to the system of Ref.~\cite{Testard2014} is tentatively related to the polymer connectivity of the microgels, which facilitates the merging of close clusters.

In this contribution we investigate whether the scaling behaviour found for the microgels of Ref.~\cite{Moreno2018} is a general result for other macromolecules 
or if it depends significantly on the architecture of the system.  
In order to quantify the growing length scale we introduce a density field representation of the macromolecules that removes artifacts arising from the local, fast
density fluctuations in the coarsening structure \cite{Testard2014}. 
We establish a set of scaling laws for the time dependence of the growing domain size during the coarsening, which are independent of the solvent quality parameter in 
analogy to time-temperature superposition principle. 
Domain growth in microgels shows a power law, though an overshoot is found in the late stage of the collapse for the case of diamond-like networks. This unusual behavior is related to the fast late merging of the regularly distributed nucleating centers.
Power-laws are also observed for the coarsening dynamics of collapsing flexible linear chains and star polymers, though with slightly smaller exponents imputable to a lower number of nucleation centers in the absence of a network structure. Semiflexible chains present a significantly higher exponent for the domain growth. 
We also analyze the fractal structure of the clusters of different sizes that are formed during the coarsening process, and find no differences between the flexible and semiflexible systems, concluding that the higher exponent for the growing length scale originates from a faster mass diffusion in the semiflexible chains induced by the stiffness. 
Finally we characterize the dynamic density correlations during the collapse and relate them with the growing length scale. A common power-law is found for this relation, irrespective of the solvent quality and macromolecular architecture, which is a typical signature of the critical nature of the process.

The paper is organised as follows. In Section 2 the model and the simulation details are presented. In Section 3 a deep analysis of the kinetics of the collapse for all the investigated systems is reported. In particular, the density field construction is introduced, and data for the domain growth  and the scaling of the dynamic correlations are discussed. Section 4 summarizes our conclusions.

\section{Model and simulation details}\label{sec2}

We performed NVT molecular dynamics simulations
of star polymers with 3 and 12 arms, semiflexible and flexible linear chains, and migrogels with a disordered and a diamond-like network structure. 
We used the bead-spring model
of Kremer and Grest \cite{Kremer1990} for the interactions. All the non-bonded interactions were implemented by a, purely repulsive, 
Weeks-Chandler-Andersen (WCA) potential \cite{Weeks1971} that was modified in order to tune the quality of the solvent. This was achieved by introducing an attractive tail, regulated by a solvent quality parameter $\phi$ that sets the solvophobicity of the monomers \cite{soddemann2001EPJE,LoVerso2015,Gnan2017,Moreno2018}. Therefore this parameter plays the role of an effective temperature in the model. The non-bonded interactions were given by:
\begin{equation}
V_{\rm nb}(r) = \left\{
\begin{array}{lll}
V_{\rm LJ}(r)=  4\epsilon\left[\left(\frac{\sigma}{r}\right)^{12}
- \left(\frac{\sigma}{r}\right)^{6} +\frac{1}{4} \right] - \epsilon\phi &\ &\  r\leq 2^{1/6}\sigma \\
V_{\phi}(r) = \frac{1}{2} \phi\epsilon \left[\cos\left(\alpha (\frac{r}{\sigma})^2 +\beta\right)-1\right]  &\ &\  2^{1/6}\sigma < r\leq 1.5\sigma \\
 0 &\ &\ r > 1.5\sigma
\end{array}
 \right.
\label{eq:vnb}
\end{equation}
The values for the parameters  $\alpha = \pi(2.25 -2^{1/3})^{-1}$ and $\beta = 2\pi -2.25\alpha$ are chosen in order to satisfy the condition that the non-bonded potential and its first derivative are continuous both at $r=2^{1/6}\sigma$ and at the cutoff $r_{\rm c}=1.5\sigma$ \cite{soddemann2001EPJE}. 
For the case $\phi=0$ the purely repulsive WCA potential is recovered and the system is in good solvent conditions.  
The quality of the solvent is worsened by choosing $\phi > 0$. Bad solvent conditions are reached when $\phi$ is higher than some critical value and the system collapses. In all the studied systems the collapse transition ($\theta$-point) occurred at $\phi \sim 0.6$ (this was estimated from the maximum of the derivative of the radius o gyration vs. $\phi$).

In addition, bonded monomers interact via a finite extensible nonlinear elastic (FENE) potential, which implements the molecular connectivity. The FENE potential reads \cite{Kremer1990}:
\begin{equation}V_{\rm FENE}(r) = -\epsilon K_{\rm F}R_0^2 \ln\left[1-\left(\frac{r}{R_0\sigma}\right)^2 \right] ,
\label{eq:fene}
\end{equation}
where $K_{\rm F} = 30$ is the spring constant and $R_0 = 1.5$ is the maximum elongation. The sum of the FENE and the non-bonded potential provides a total interaction between two connected monomers showing a deep minimum at $r \sim 0.95$, which guarantees uncrossability and prevents violation of the topological constraints.
In the case of the semiflexible linear chains the bending stiffness was implemented through the
worm-like model \cite{Zierenberg2016, Kratky1949}.  Thus, the interaction for the polymer bending has the form:
\begin{equation} V_{\rm bending}(\theta) = K_{\rm s}(1-\cos\theta) ,
\label{eq:bending}
\end{equation}
where $\theta$ is the angle between two successive bonds and $K_{\rm s}$ is the strength of the bending. We used a value $K_{\rm s} = 5$.
The units of energy, length, mass and time are respectively $\epsilon$, $\sigma$, $m$ and $\tau=(\sigma^2 m/\epsilon)^{1/2}$ where $m$ is the mass of a monomer.
In the rest of the paper all the numerical values will be given in reduced units $\epsilon = \sigma = m = \tau = 1$.

We performed MD simulations at temperature $T = \epsilon/k_{\rm B} = 1.0$ (with $k_{\rm B}$ the Boltzmann constant) using a Langevin
thermostat \cite{Izaguirre2001,Smith2009}. Thus, the force experienced by the monomers is:
\begin{equation}\ddot{\textbf{r}}_{i} =  -{\bf \nabla} V(\textbf{r}_i) - \gamma \dot{\textbf{r}}_{i} + \sqrt{2 \gamma k_{\rm B} T} {\bf \zeta}(t)
,
\label{eq:langevin}
\end{equation}
where $\textbf{r}_i$ is the position vector and $V(\textbf{r}_i)$ is the total interaction potential for the monomer of index $i$. The second term of the right side of Equation~(\ref{eq:langevin}) represents viscous damping, with $\gamma$ the friction coefficient. The last term is a random uncorrelated force, $\langle \zeta^{\alpha}_i(t)\zeta^{\beta}_j(t')\rangle = \delta_{i,j}\delta_{\alpha,\beta}\delta(t-t')$ (with $\alpha,\beta$ the Cartesian components), representing the collisions with solvent particles. The Langevin thermostat acts therefore as an implicit solvent, in which every particle interacts independently with the solvent, but hydrodynamic interactions between solute particles are not considered. Their inclusion would require the use
of e.g., Lattice Boltzmann \cite{latticeboltzmann}, multi-particle collision dynamics (MPCD) \cite{Malevanets1999} or 
dissipative particle dynamics \cite{Espanol1995} methods, involving a huge computational cost due to the
size of the investigated systems and the big boxes needed to avoid finite size effects. For example, in the widely used MPCD method, radii of gyration $R_{\rm g} \sim 50$ would require using boxes
of side $L_{\rm box} \sim 200$ and at least $5L_{\rm box}^3 = 4 \times 10^7$ solvent particles for both correctly implementing the hydrodynamic interactions 
and avoiding significant finite-size effects \cite{Mussawisade2005}.
Still, based on previous evidence \cite{Pham2008,Nikolov2018,Camerin2018,Keidel2018} we do not expect that hydrodynamics will lead to {\it qualitatively} different results from those presented here.
We used a time step $\delta t = 0.005$, and  a friction $\gamma = 0.05$, which is high enough for good thermalization and low enough to prevent strong damping that would slow down the dynamics to time scales requiring a huge computational cost.

After generating them, equilibration of all the investigated systems (flexible and semiflexible chains, stars, disordered and diamond microgels) was performed in the limit of good solvent ($\phi=0$). To investigate the coarsening kinetics, the systems equilibrated at $\phi =0$ were quenched at infinite rate to different values of the solvent parameter $\phi$ well below the $\theta$-point, and coarsening was analyzed from the quenching instant ($t=0$) until the macromolecule reached its equilibrium fully collapsed state.

The linear chains were constructed as simple strings of beads and springs (adding the bending interaction in the case of the semiflexible chains).
We used $N=1600$ and 400 beads for the flexible and semiflexible chain, respectively.
The star polymers were constructed by linking flexible linear arms of 800 beads to a central bead. 
Several protocols have been recently proposed to generate realistic models of microgels \cite{Gnan2017,Moreno2018,Minina2019,Ninarello2019,Rudyak2019,Rovigatti2019}
beyond the regular networks usually employed in the literature.
In our method we take inspiration from the synthesis of microgels in microfluidic cavities. This experimental route takes place via confinement in a droplet of  pre-existing linear polymer chains, and further inter- and intra-molecular irreversible association  \cite{Weitz2010}.
In our model each confined single chain consists of $N$ beads, and in each chain $N_{\rm r}$ of these beads are reactive (cross-linkable) groups, randomly distributed along the polymer backbone and with at least one inert group between consecutive reactive groups to prevent trivial cross-links.  
In order to implement the cavity, a confining spherical, purely repulsive, LJ potential is applied to each monomer:  
\begin{equation}
V_{\rm wall}(r_{\rm w}) = \left\{
\begin{array}{lll}
4\epsilon\left[\left(\frac{\sigma}{r_{\rm w}}\right)^{12}
- \left(\frac{\sigma}{r_{\rm w}}\right)^{6} +\frac{1}{4} \right] &\ &\  r_{\rm w} \leq 2^{1/6}\sigma \\
0 &\ &\ r_{\rm w} > 2^{1/6}\sigma
\end{array}
\right. 
\end{equation}
where $r_{\rm w}$ represents the shortest distance from the monomer to the spherical wall. 

We performed the cross-linking of $N_{\rm cha}=36$ chains in a cavity of radius $R_{\rm cav} =55$.
Each chain had $N=600$ beads, so that the finally generated microgel had 21600 beads. 
The number of reactive beads in each chain was $N_{\rm r}=12$, 
i.e., the fraction of cross-linked monomers in the microgel was $f = N_{\rm r}/N = 0.02$.
The number density used in the synthesis was $3N_{\rm cha}N/(4 \pi R^3_{\rm cav}) \approx  0.03$, which qualitatively
corresponds to experimental concentrations of about 30 mg/mL \cite{Kremer1990,Moreno2016JPCL,gonzalezburgos2018}.
After equilibration of the chains inside the cavity the cross-linking of the reactive groups was activated.
A permanent bond (modeled by the FENE potential) between two reactive groups was formed if: 
(i) none of them was already bonded to another reactive group, and (ii) they were at a mutual capture distance $r < 1.3\sigma$. 
A random choice was made in case of multiple candidates within the capture distance.
To speed up the late stage of the cross-linking process ($\leq 6 $ remaining unbonded reactive groups), a random pair was chosen from the unreacted groups 
and an attractive harmonic interaction between the constituents of the pair was implemented, in order to approach them to the capture distance and form the bond.
After forming the bond the microgel was equilibrated and the procedure was repeated until full completion of the cross-linking. Then the cavity was removed to allow for swelling and equilibration of the obtained microgel. Cross-linking of 50 initial realizations, with the same former values 
of $N$, $N_{\rm cha}$, $N_{\rm r}$ and  $R_{\rm cav}$ was perfomed, leading to microgels with the same number of monomers
and cross-links but topologically polydisperse \cite{Moreno2018}. 
It was found that a large number of the cross-links (about 65 \%) occurred between reactive groups belonging to the same polymer chain, 
forming loops. These kind of cross-links do not contribute to the
connectivity of the network and are elastically inactive. On the other hand in the diamond-like microgels no loops are present and all the cross-links 
are elastically active. Therefore, for a fair comparison with the disordered ones, the diamond-like microgels were constructed with the same fraction
of cross-links (nodes) as the average number of intermolecular bonds in the disordered microgels, tuning the number of nodes so that the total mass
of both kinds of microgels was essentially the same \cite{Moreno2018}. Thus, we simulated a diamond-like microgel of $N=21615$ beads containing 78 nodes.
The diamond-microgel was generated by placing the cross-links in the nodes of a regular diamond network, and by connecting every pair of nearest-neighbour nodes through linear bead-spring rods \cite{olvera2011,winkler2014,winkler2016,Ahualli2017,Sean2018}. All the beads out of a sphere containing the selected $N$ beads were removed.

We computed the time-averaged asphericity $a$ \cite{rawdon2008} of each disordered microgel in the swollen state ($\phi=0$)
and obtained the corresponding distribution $P(a)$. For the analysis of the coarsening kinetics
we selected three disordered microgels, at the center and at the two extremes of the distribution of asphericities. 
The corresponding radius of gyration, at $\phi=0$, of the selected disordered microgels is $R_{\rm g}= 49.3, 52.4$ and 64.0 for $a=0.02, 0.06$ and 0.14, respectively. In the rest of the paper these disordered microgels will be denoted in the figure legends as I-II-III from lower to higher asphericity.
The size of the other investigated systems at $\phi=0$ is  $R_{\rm g}=51.8$ (diamond-like microgel), 45.3 (3-arm star), 48.9 (12-arm star), 39.9 (flexible chain) and 26.4 (semiflexible chain).

\section{Coarsening Kinetics}\label{sec3}

The coarsening kinetics is characterized by a growing length scale. The quantitative characterization of such a length scale
can be easily affected by artifacts originating from the `structural noise' emerging during the macromolecular collapse (bridges that connect the clusters of monomers, small halls, protrusions, etc). To avoid these artifacts we introduce a smooth representation of the macromolecules through a coarse-grained density field. This method is based, originally, on the characterization of the growing length scales 
in a coarsening binary Ising system \cite{Majumder2011} and was later applied in the continuous space to 
a liquid-gas phase separating system \cite{Testard2011,Testard2014}. In the first case the thermal noise effects on the coarsened structure were removed by using a majority spin rule, i.e. by replacing each spin by the majority spin of its nearest neighbours. In the second case, the real particles were substituted by their local densities averaged over their nearest environment. 
These averaging procedures smooth the interface corrugations, fill the smallest holes, and delete the smallest clusters in the coarsening structure. In this way the smooth density field avoids, in the calculation of the growing length scale, the effect of non-relevant minimal paths or artificial interruptions of long paths within the dense domains. 
In our systems we construct our coarse-grained density field as follows:
\\
i) we divide the space into cubic cells of side $\delta$;
\\
ii) we define the local density (for each cell) by the number of monomers in a sphere:

\begin{equation}
\rho({\bf r}) = 3n({\bf r})/(4\pi r_{\rm c}^{3})
\end{equation}
where $r_{\rm c}$ is the cutoff radius of the sphere and $n({\bf r})$ is the
number of monomers at a distance $d$  $\leq$  $r_{\rm c}$ from the position $\bf r$ of the cell center;
\\
iii) we fix the value of the coarse-grained density
at ${\bf r}$ as a weighted average of the local density over the surrounding cells:
\begin{equation}
\bar{\rho}({\bf r}) = \frac{1}{8} \left[ 2\rho({\bf r}) +\sum_{\bf k}\rho({\bf r} + \delta{\bf k}) \right]
\end{equation}
where the sum is performed over the directions ${\bf k} \in \{(\pm 1,0,0),(0,\pm 1,0),(0,0,\pm 1)\}$.
The grid size $\delta$ and the cutoff radius $r_{\rm c}$ are chosen 
in order to get a smooth density field while keeping sufficient spatial resolution in the representation of the real system.
This is achieved by using values $\delta \sim 0.5$ and $r_{\rm c} \sim 1$. In what follows we will present results for
$\delta = 0.5$ and for two specific values  $r_{\rm c} = 1.0$ and $r_{\rm c} = 1.2$.
By using a threshold value $\rho_{\rm min}$ in the density field construction the macromolecules can be seen as a coarsening biphasic system composed by 'empty' and 'filled' domains. These domains are identified
according to their low ($\bar{\rho}({\bf r}) \le \rho_{\rm min}$) or high ($\bar{\rho}({\bf r}) > \rho_{\rm min}$) local density, respectively.

\begin{figure}[pt!]
\centering
\includegraphics[width=1.0\textwidth]{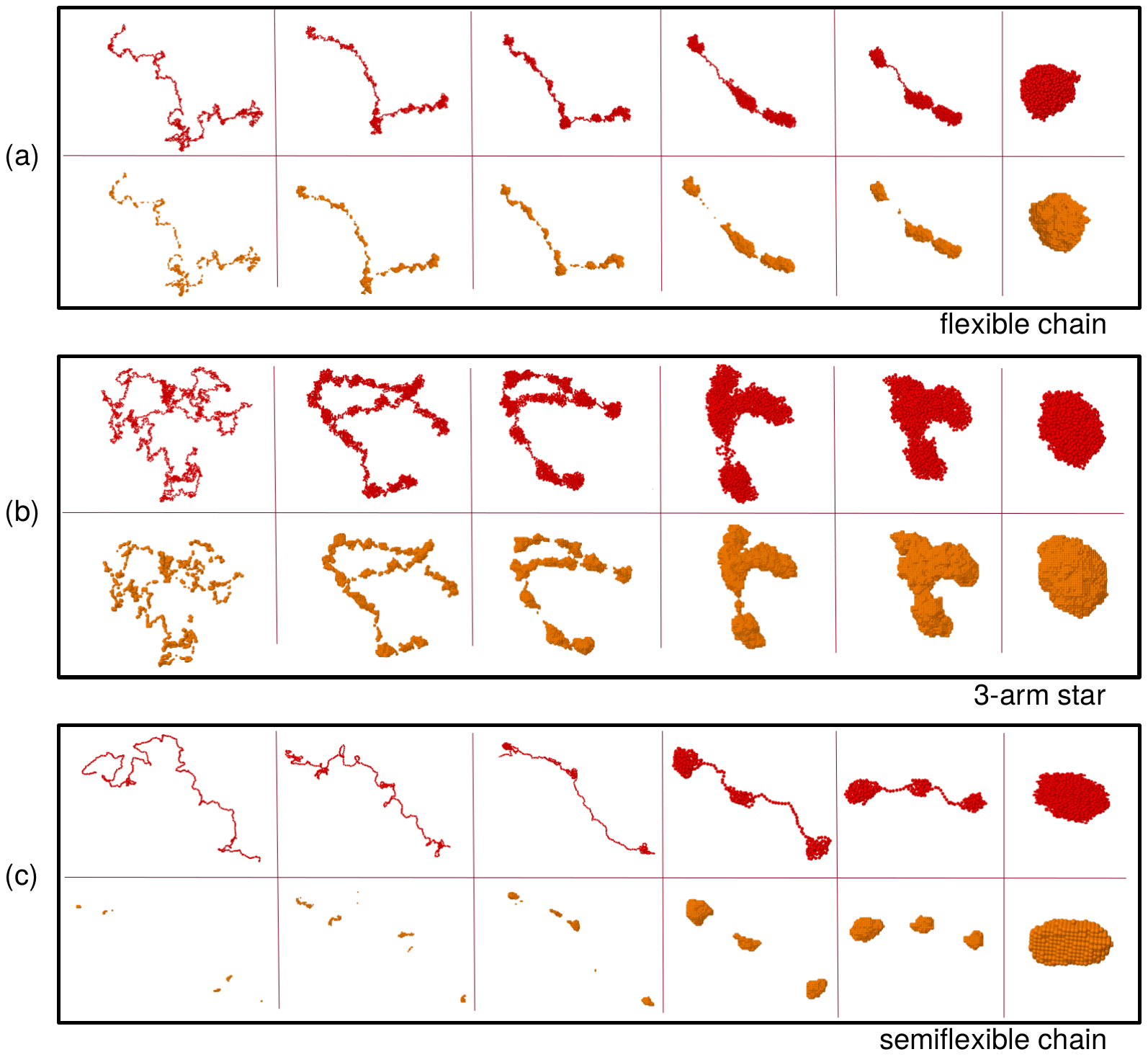}
\caption{Snapshots of real (red beads) and density field (orange beads) coordinates of the flexible linear chain
(a), the 3-arm star (b) and the semiflexible linear chain (c). In all cases the collapse occurs at solvent parameter
$\phi=1.2$, and the parameters $r_{\rm c} = 1.0$ and $\rho_{\rm min} = 0.6$ are used to construct the density field. 
The times for each panel are, from left to right:
$t = 5, 50, 126, 251, 316, 1000$ ((a) and (b)), and $t = 5, 50, 126, 398, 500, 8000$ (c).
}
\label{fig:snaps-field} 
\end{figure}

Figure~\ref{fig:snaps-field}
shows snapshots of the real-coordinates (read beads) and the density field representations (orange beads) at different times during the collapse of a flexible chain (a), a 3-arm star (b) and a semiflexible chain (c), respectively. 
In the real-coordinate representation all the momomers are displayed. In the density-field representation only the filled cells ($\bar{\rho}({\bf r}) > \rho_{\rm min}$) are shown.
All these snapshots  correspond
to $\phi=1.2$, a bad-solvent state well below the $\theta$-point. The final state reached at the end of the simulation (last columns on the right) is in all cases a fully collapsed macromolecule.
The collapse experienced by the flexible chain and the star polymer (panels (a) and (b)) begins with the formation of clusters of monomers along the chain, then those merge by withdrawing monomers from the bridges connecting them, in a longitudinal diffusion process. 
In the stars this process also includes merging of clusters of different arms.
The collapse for the semiflexible chain (panel (c)) seems to be qualitatively different: the dense regions are better defined, as it becomes evident in the density field representation. At early times, when the conformations are close to those of $\phi=0$, there are many less clusters than in the flexible case. Indeed small fluctuations leading to local transient clustering are strongly hindered by bending stiffness. However, as time goes on and the effective monomer attraction starts to drive the collapse, the clusters quickly grow up before starting to merge into larger clusters. In contrast to the flexible systems where the chain or arm backbone can
still perform broad lateral fluctuations at early and intermediate times, in the semiflexible chains merging of the cluster proceeds along a quasi-rodlike structure during the whole process.
In the next subsections we quantify the observed similarities and differences, by analyzing the domain length distribution, the domain growth rate, as well as the dynamic correlations, during the coarsening process.

\subsection{Chord length distributions and domain growth}

The construction of the coarse-grained density field allows  to measure the distribution of the domain size from the obtained smooth biphasic structure.
We define an `interfacial cell' \cite{Moreno2018} in our system  as a filled cell with at least one adjacent empty cell.
In order to compute the characteristic lengths of the coarsened structures we use the definition of chord \cite{Testard2011,Testard2014}. This is a straight path, along one of the three $x,y,z$-directions of the grid, which is formed just by filled cells and whose two end cells are interfacial cells.
To compute the distribution of chord lengths of a given macromolecular configuration at a given time and solvent parameter $\phi$, we sampled all the existing chords by following all the possible paths along the three directions within a volume containing all the filled cells.
To improve statistics 5 random rotations of the former configuration were taken, 
and the whole procedure was repeated over 5 independent realizations of the same macromolecule.

\begin{figure}[t!]
\centering
\includegraphics[width=.95\textwidth]{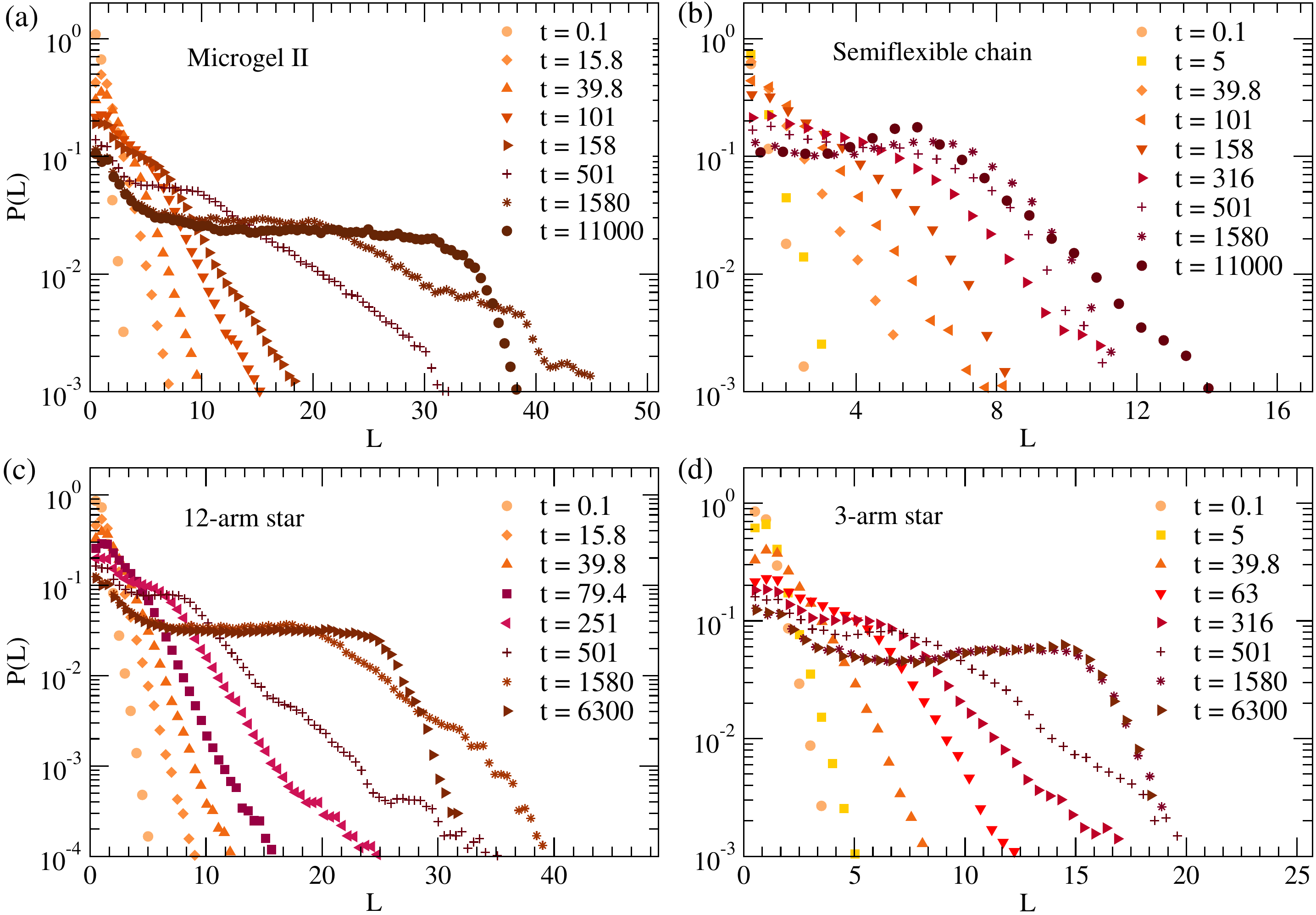}\hspace{0.3cm}
\caption{Normalized distribution of chord lengths at $\phi=1.2$ and different times during the collapse of the
disordered microgel of middle asphericity (a), the semiflexible chain (b), 12-arm star  (c) and 3-arm star  (d). The chord lengths are calculated for a density field
with $r_{\rm c}=1.0$ and $\rho_{\rm min}=0.6$.}
\label{fig:distlength} 
\end{figure}

Figure \ref{fig:distlength} shows the normalized distributions of chord lengths, $P(L)$, at $\phi=1.2$ and different times during the collapse of the disordered microgel of middle asphericity (Figure \ref{fig:distlength}a), the semiflexible chain (\ref{fig:distlength}b), 12-arm star polymer (\ref{fig:distlength}c) and 3-arm star polymer (\ref{fig:distlength}d). 
$P(L)$ shows the characteristic length distribution observed in coarsening systems  \cite{Levitz1998,Atsuko2003,Majumder2011,MajumderEPL2011,Testard2014,Majumder2017}. In particular, for earlier times $P(L)$ shows an exponential decay and extends over longer distances
as time increases, which is a consequence of the growth of the filled domains during the coarsening. 
As expected, the exponential behavior saturates at long times when the fully collapsed state is reached. The flat plateau originates
from the equiprobable different straight paths that connect two points, at both sides of the outer interface, in the
fully collapsed state (where empty cells are absent). The drop from the plateau obviously reflects the finite size of the collapsed object .
Interestingly, the $P(L)$ of the semiflexible chain shows an approximate self-similar behavior over time, i.e., the decay is shifted to longer times but shows roughly the same shape, in contrast with the other systems for which the slope of the exponential is strongly time-dependent.
This is consistent with the growth and transport process of the nucleation centers anticipated in Figure \ref{fig:snaps-field}c. In fact, as shown there, the clusters are better defined and further apart, and they grow considerably before coalescing in a single cluster. Instead, in the flexible systems (panels
(a,b) of Figure \ref{fig:snaps-field}) the clusters grow and merge in a much more gradual way until the single globule is formed at late times. 

\begin{figure}[t!]
\centering
\includegraphics[width=.95\textwidth]{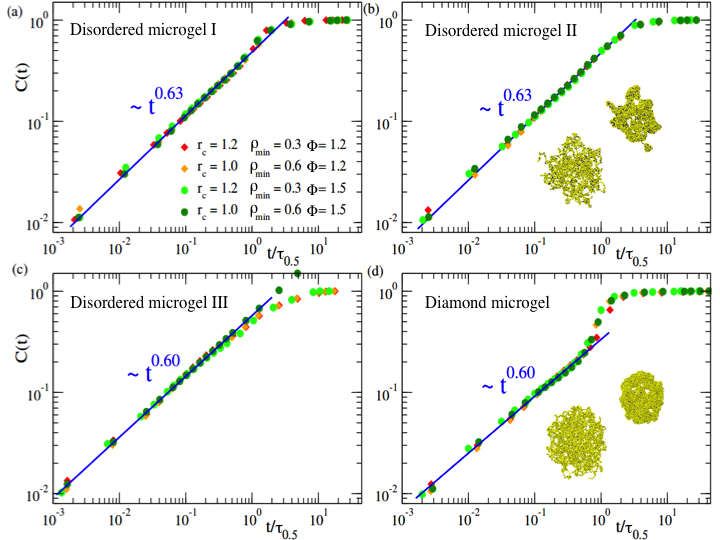}
\caption{Relative domain size $C(t)$ over time during
the collapse at $\phi = 1.2$ and $\phi = 1.5$ of three disordered microgels of small (a),
middle (b) and large asphericity (c) and of a diamond microgel (d). Different data sets correspond to different values of the parameters $r_{\rm c}$
and $\rho_{\rm min}$ used to construct the density field. The time $t$ is normalized by $\tau_{0.5}$, 
defined as the time when $C = 0.5$. Straight lines in all panels are fits to a
power-law time dependence. Exponents are indicated. In panels (b) and (d) we show typical simulation snapshots at early and intermediate times.}
\label{fig:ct-densfield-microgel} 
\end{figure}

We use the information on the former distributions to quantify the growing length scale characteristic of the coarsening process.
The mean domain length for a given time $t$ is obtained from the first momentum of the distribution $P$ at time $t$ , i.e.,
$L(t) = \int L' P(L'; t) dL'$.
We find the same qualitative behavior of $P(L)$ for several choices of $r_{\rm c}$ and $\rho_{\rm min}$, whereas quantitatively $L(t)$ depends on the specific 
parameters used to construct the density field. Indeed, choosing, e.g.,  a lower value for $\rho_{\rm min}$ implies having a higher number of filled cells in the density field, and consequently a larger value of $L$. For this reason we quantify the domain growth rate in terms of the relative domain size, defined as:
 \begin{equation}
C(t) = \dfrac{L(t)-L(0)}{L(\infty)-L(0)}
 \end{equation}
In this way $C(t)$ represents a normalized mean chord length growing from zero at $t=0$ to 1 at late times in the collapsed state. 
Figure \ref{fig:ct-densfield-microgel}  shows the domain growth for three different topologies of the disordered microgels (corresponding to a low, middle, and large value in the distribution of the asphericity parameter), in comparison with the diamond-like network. For each system the panel includes data of $C(t)$ for several selections of the parameters $r_{\rm c}$ and $\rho_{\rm min}$. The absolute times have been rescaled by the time $\tau_{0.5}$, defined as $C(\tau_{0.5}) =0.5$. As can be seen for all systems, after this time rescaling the different data sets of $C(t)$ nicely overlap. This demonstrates that the density field approach is consistent, as it provides a time dependence of $C(t)$ that is independent of any reasonable choice of the parameters defining the density field.
Furthermore a good overlap is observed also for different values of the solvent quality parameter ($\phi = 1.2$ and $\phi = 1.5$), indicating that the coarsening kinetics follows an effective time-temperature superposition principle.

As can be seen in Figure \ref{fig:ct-densfield-microgel}, the function $C(t)$ reveals that the coarsening length scale grows by following a sublinear power law $\sim t^x$. The exponents are $x \gtrsim 0.6$ for all the microgels, irrespective of the microstructural degree of disorder. The exponents obtained here for $f=0.02$ are slightly smaller than those found in our previous work \cite{Moreno2018} ($x \sim 0.7$) for a much higher degree of cross-linking ($f=0.1$). 
We believe that the slightly higher value of the exponent for $f=0.1$ is just an artifact originating from the proximity of the cross-links
in such a system. Namely, since the higher density of monomers around a cross-link enhances its propensity to become a nucleating center, the much smaller distance between cross-links in the case $f=0.1$ than in $f=0.02$ may effectively accelerate the coalescence of the growing clusters, 
leading to the observed faster domain growth.
  
Our previous work in microgels with higher $f$ suggested that the diamond network exhibited the same power-law in $C(t)$ that the disordered networks. However, for the diamond network this behaviour was developed only in a narrow time window and the results were not conclusive \cite{Moreno2018}.
The results in Figure \ref{fig:ct-densfield-microgel}d for $f=0.02$ clearly confirm the power law regime over 3 time decades.
These data also confirm the acceleration of the domain growth in the late stage of the collapse of the diamond network \cite{Moreno2018}, which is at most a marginal effect
in the disordered networks, where the power-law continues and is a good approximation until the collapsed state is reached and the ultimate plateau emerges. The strong late acceleration in the diamond network is tentatively related to the regular spatial distribution of the cross-links acting as preferential nucleating centers,
which leads to a more homogeneous collapse until all holes in the structure vanish in a short time window (compare snapshots in panels (b) and (d)),
and the filled paths in the density field experience a late sudden growth.

\begin{figure}[t!]
\centering
\includegraphics[width=.9\textwidth]{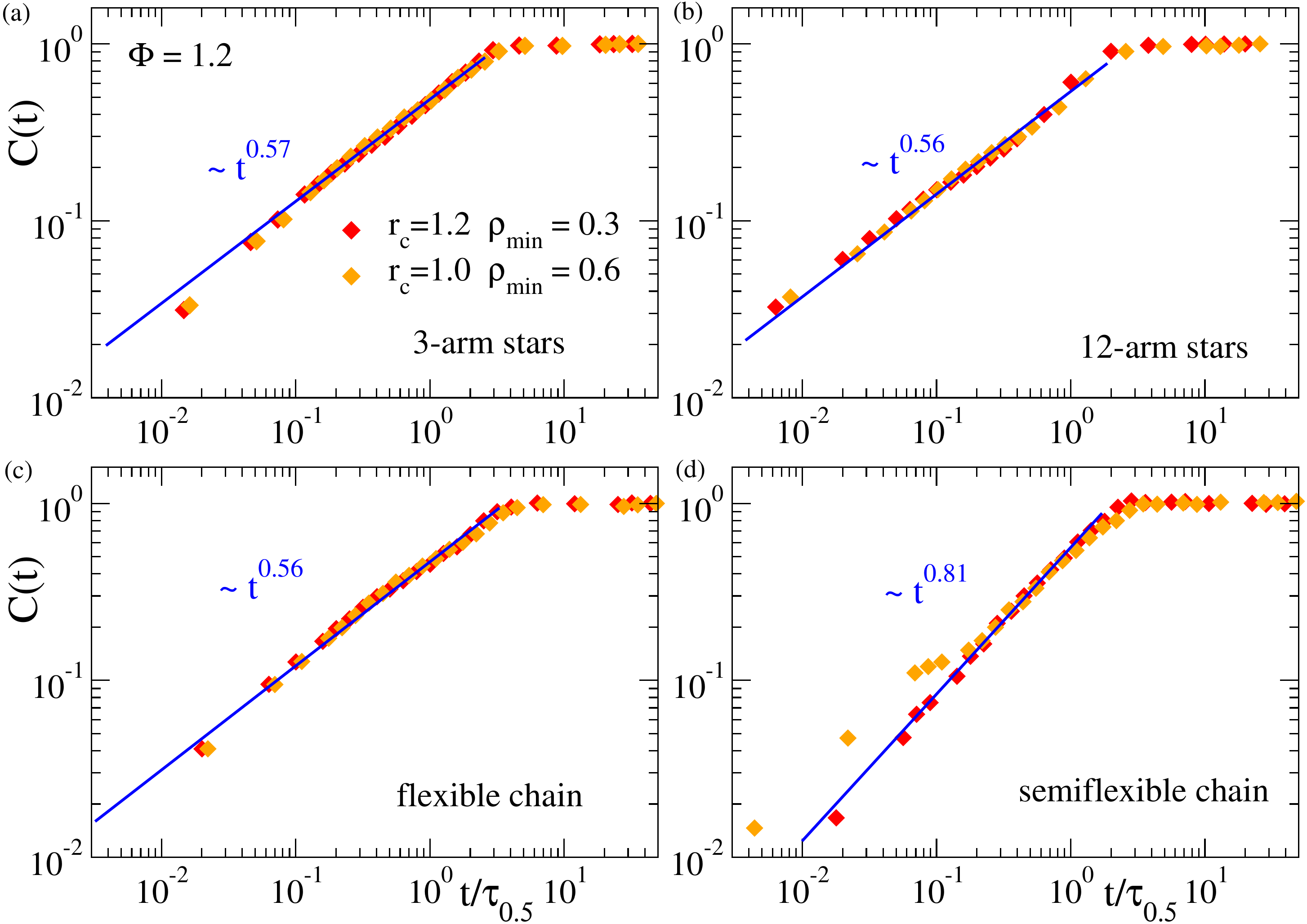}\hspace{0.01cm}
\vspace{0.1cm}
\caption{Relative domain size $C(t)$ over time during
the collapse at $\phi = 1.2$ of 3-arm stars (a),
12-arm stars (b), flexible chains (c) and semiflexible chains (d). Different data sets correspond to different values of the parameters $r_{\rm c}$
and $\rho_{\rm min}$ used to construct the density field. 
The time $t$ is normalized by $\tau_{0.5}$, defined as in Figure~\ref{fig:ct-densfield-microgel}. Straight lines in all panels are fits to
a power-law time dependence (exponents are indicated). 
}
\label{fig:ct-densfield-total} 
\end{figure}

The results presented above confirm the common scaling law $C(t) \sim t^{0.6}$ for the domain growth in collapsing microgels, irrespective of their topology (disordered or regular networks). In order to search for a more general scenario, we investigate the scaling properties of the coarsening kinetics in other very different architectures, as stars with different number of arms, and flexible and semiflexible linear chains. We follow the same procedure as for the microgels, constructing the density field and analyzing the domain growth in this representation of the macromolecule. Figure~\ref{fig:ct-densfield-total} shows the corresponding results of $C(t)$ for the stars and linear chains.
We find that the systems without bending stiffness (the stars and the flexible linear chains) essentially show the same power-law 
$C(t) \sim t^{0.56}$, with an exponent that is just slightly smaller than those found for the microgels. The only investigated system with bending stiffness, i.e., the semiflexible linear chains, shows a different power-law, with a clearly higher exponent $C(t) \sim t^{0.8}$.
In a first approximation it is possible to venture that the flexible nanoparticles as microgels and star polymers act in a similar fashion during most of the coarsening process: at distances shorter than the arm length and the mesh size, the arms and strands behave like linear flexible chains. Microgels show an exponent of $\sim 0.6$ for the domain growth, slightly larger than the value $\sim 0.56$ found for stars and single flexible chains. This difference might be connected with the higher concentration of nucleating centers in the microgels (due to the presence of the cross-links). Comparatively, at the local scale the semiflexible chain is influenced by the presence of the bending, which reduces its lateral fluctuations.
This reduction may promote a faster and more homogeneous aggregation of mass along the chain contour, leading to the observed steeper growth of the coarsening length scale.

\subsection{Cluster analysis}

\begin{figure}[t!]
\centering
\includegraphics[width=.6\textwidth]{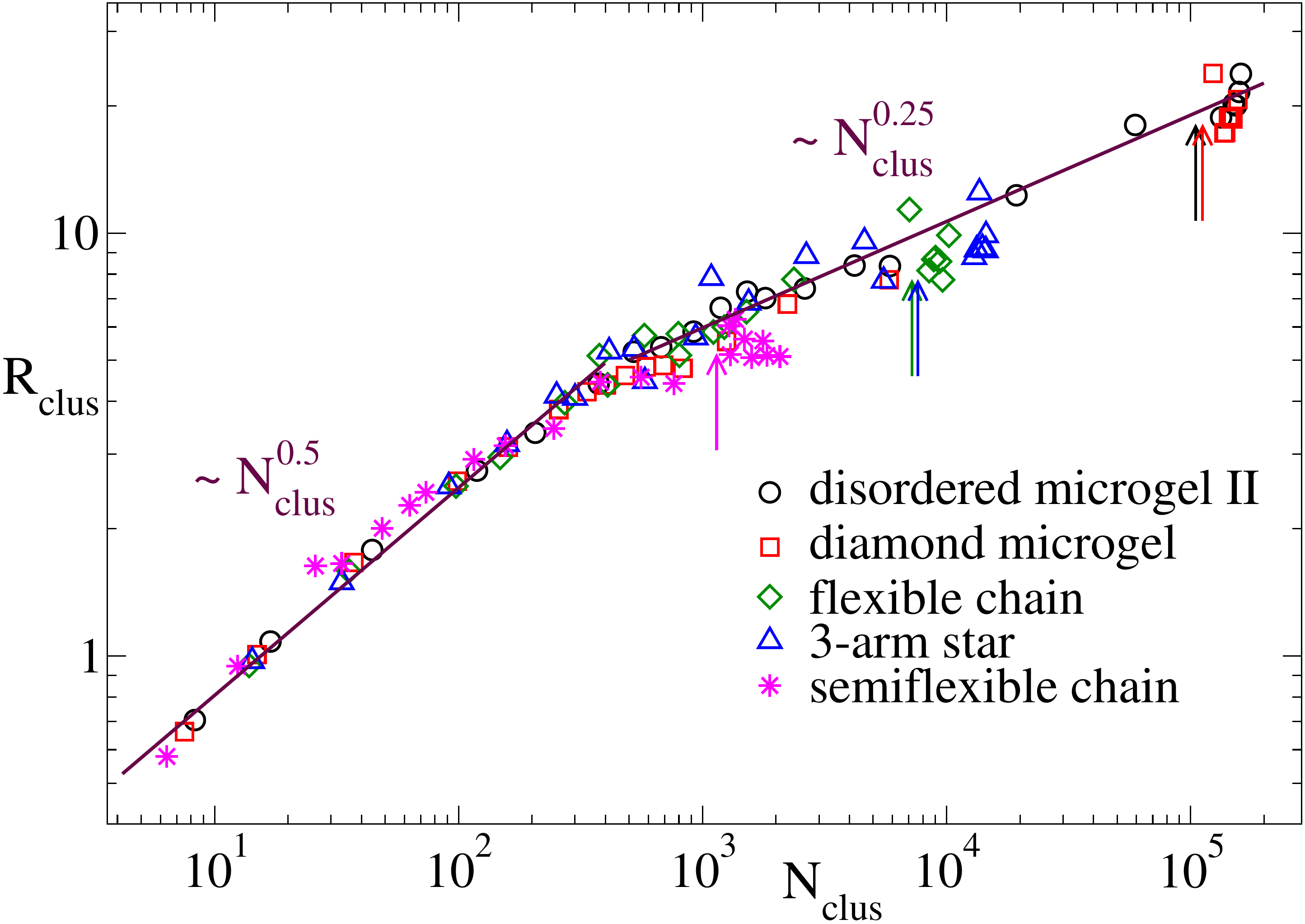}
\caption{Mean cluster radius vs. mean cluster population at $\phi = 1.2$, for clusters of filled cells
in the density field representation with parameters $r_{\rm c} = 1.0$ and $\rho_{\rm min} = 0.6$.
Data are shown for the disordered microgel of middle asphericity, the diamond network, the 3-arm star and the
flexible and semiflexible linear chains. The arrows indicate the approximate cluster population 
at the saturation point prior the formation of the late plateau in  $N_{\rm clus}(t)$.
Lines are power-laws, exponents are indicated.}
\label{fig:n-r-clusfield} 
\end{figure}

In this subsection we shed some ligth on the microscopic origin of the different scaling exponents for the
domain growth in the flexible systems and the semiflexible linear chains. We first analyze 
whether stiffness plays a role on the conformational properties of the dense regions 
that progressively emerge in the coarsening structure. Namely we analyze their fractal behaviour, 
i.e., the power-law dependence between the size and the mass of such dense regions.
This can be done
by defining clusters of filled cells ($\rho > \rho_{\rm min}$) in the density field representation, 
and determining the relation between the mean radius $R_{\rm clus}$ and mean population $N_{\rm clus}$ of the clusters. 
Two filled cells
belong to a same cluster if they are adjacent, i.e., if they are connected by a vector
$\delta{\bf k} \in \{(\pm \delta,0,0),(0,\pm \delta,0),(0,0,\pm \delta)\}$.
The mean radius $R_{\rm clus}$ is just obtained as $\langle R_{\rm g}^2 \rangle^{1/2}$, with $R_{\rm g}$ the
radius of gyration of the cluster. At each time we average the former quantities over all the clusters,
obtaining the time-dependence $R_{\rm clus}(t)$ and  $N_{\rm clus}(t)$. Substitution of time provides a univoque relation $R_{\rm clus}(N_{\rm clus})$
for the fractal behavior of the clusters. Figure~\ref{fig:n-r-clusfield} shows $R_{\rm clus}(N_{\rm clus})$
for several representative flexible systems (see legend) and for the semiflexible chains. All the systems show two 
different power-laws $R_{\rm clus} \sim N_{\rm clus}^\nu$ at both sides of a crossover value $N^{\ast} \sim 400$. For small clusters ($N_{\rm clus} < N^{\ast}$) an exponent $\nu \approx 0.5$ is found, which indicates approximate Gaussian statistics
of the cluster conformations. Big clusters, which are formed in the late stage of the coarsening process, 
show an exponent $\nu \approx 0.25$, reflecting the expected compact structures that progressively merge until reaching the equilibrium globule.

\begin{figure}[t!]
\centering
\includegraphics[width=.9\textwidth]{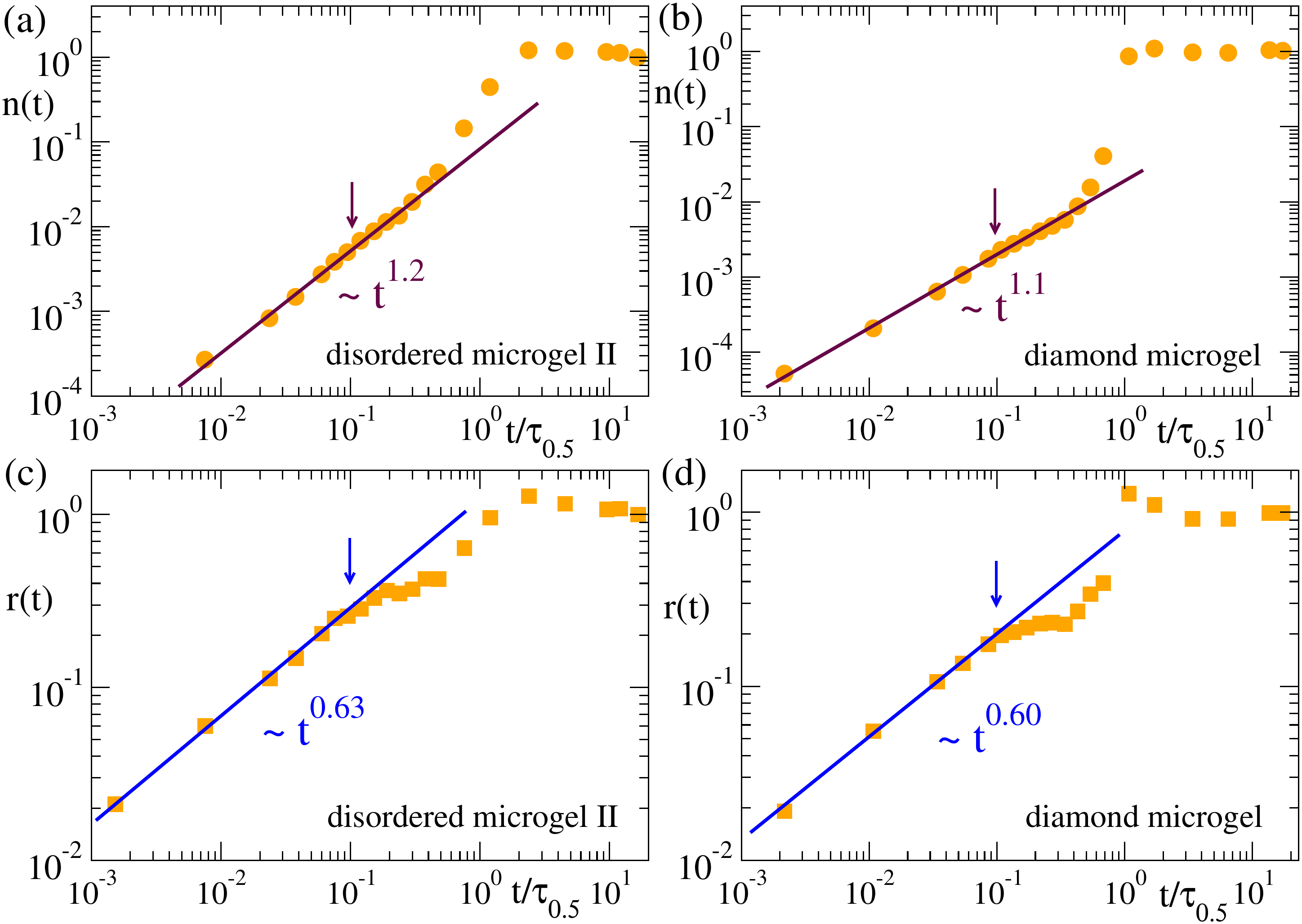}
\caption{Normalized population (a,b) and radius (c,d) of the clusters in the density field representation of the 
disordered microgel with middle asphericity (a,c) and the diamond microgel (b,d). Data correspond to a solvent quality 
parameter $\phi= 1.2$, and a density field representation with parameters $r_{\rm c} = 1.0$ and $\rho_{\rm min} = 0.6$.
The arrows indicate the time scale for which $N_{\rm clus}(t) = N^{\ast} \approx 400$.
Times are rescaled by $\tau_{0.5}$ as defined in Figure~\ref{fig:ct-densfield-microgel}.}
\label{fig:nt-rt-clusfield-micro} 
\end{figure}
\begin{figure}[t!]
\centering
\includegraphics[width=.9\textwidth]{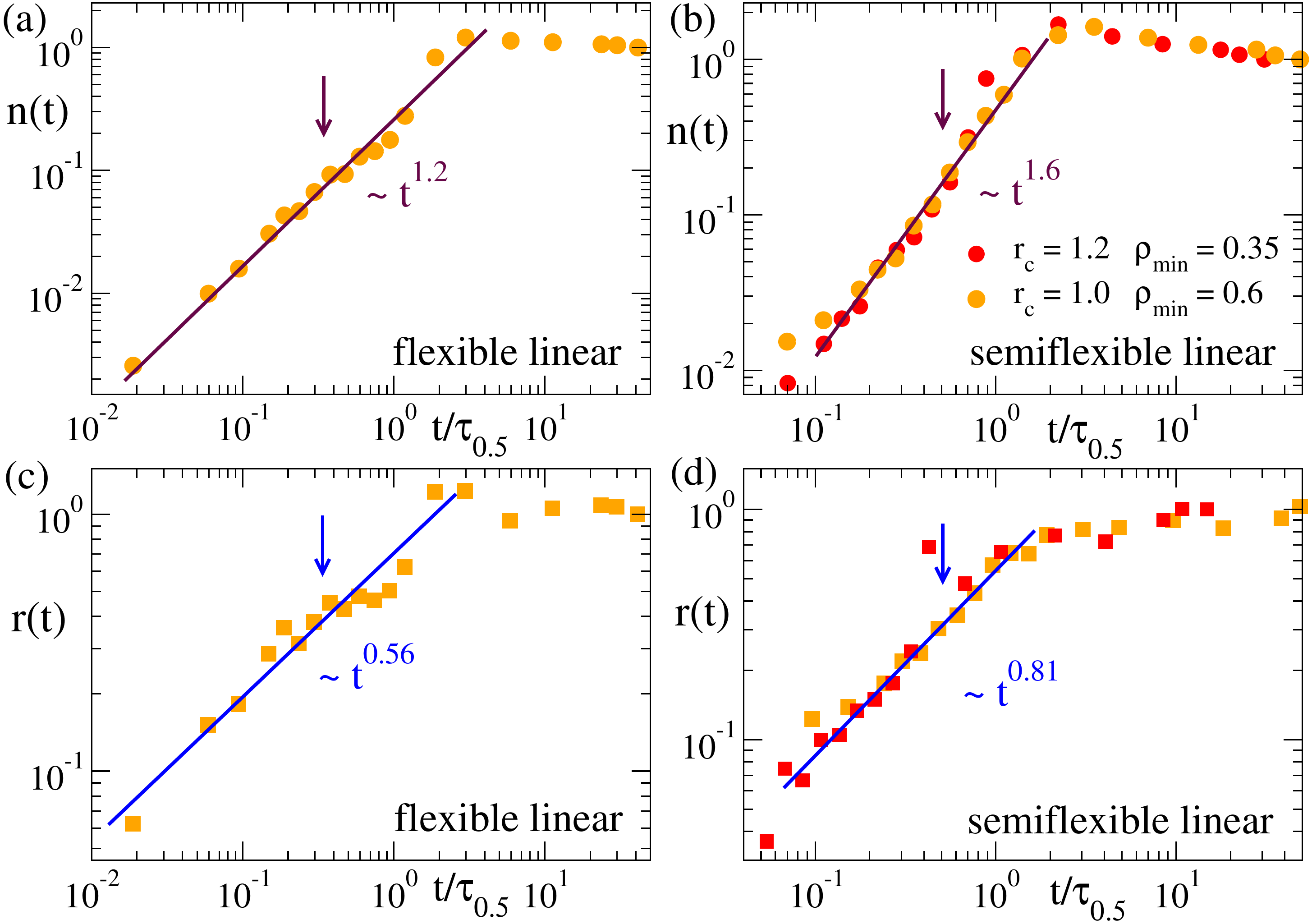}
\caption{Normalized population (a,b) and radius (c,d) of the clusters in the density field representation of the 
flexible (a,c) and semiflexible linear chain (b,d). Data correspond to a solvent quality 
parameter $\phi= 1.2$. Values ($r_{c} = 1.0$, $\rho_{\rm min} = 0.6$) and ($r_{c} = 1.2$, $\rho_{\rm min} = 0.35$) are
used to construct the density field.
The arrows indicate the time scale for which $N_{\rm clus}(t) = N^{\ast} \approx 400$.
Times are rescaled by $\tau_{0.5}$ as defined in Figure~\ref{fig:ct-densfield-microgel}.}
\label{fig:nt-rt-clusfield-linear} 
\end{figure}

Figures~\ref{fig:nt-rt-clusfield-micro} and \ref{fig:nt-rt-clusfield-linear} show the time dependence of the cluster population and size for the microgels and the flexible and semiflexible linear chains.
In analogy with the normalized function for the domain growth $C(t)$, we define the normalized population and size of the
clusters as $n(t) = (N_{\rm clus}(t)- N_{\rm clus}(0))/(N_{\rm clus}(\infty)- N_{\rm clus}(0))$ and
$r(t) = (R_{\rm clus}(t)- R_{\rm clus}(0))/(R_{\rm clus}(\infty)- R_{\rm clus}(0))$, respectively, so that both $n(t)$
and $r(t)$ grow from 0 to 1.
The arrows in all panels of Figures~\ref{fig:nt-rt-clusfield-micro} and \ref{fig:nt-rt-clusfield-linear}
indicate the time scale for which $N_{\rm clus} = N^{\ast}$.
As shown in Figure~\ref{fig:n-r-clusfield}, the clusters in all systems have the same fractal behaviour.
However, their evolution in time (Figures~\ref{fig:nt-rt-clusfield-micro} and \ref{fig:nt-rt-clusfield-linear}) can 
strongly depend on the system, namely there are clear differences between the flexible and semiflexible systems. 
In other words, coarsening leads to mass aggregation into the same kind of clusters for all the systems, but the rate of mass aggregation 
can be significantly affected by bending stiffness. 
For the time scales where the mean population of the clusters is smaller than $N^{\ast}$ (clusters are still relatively small
and approximately Gaussian), we find a power-law $n(t) \sim t^{\beta}$, with exponent $\beta \approx 1.2$ for the flexible systems, and a much higher
value $\beta = 1.6$ for the semiflexible chains. 
Not surprisingly, the growing length scale of the clusters follows a power-law 
$r(t) \sim t^{\gamma}$ that is compatible with the same exponents found in $C(t)$ for the growing domain size
(see Figures~\ref{fig:ct-densfield-microgel} and \ref{fig:ct-densfield-total})
---though clusters and domains represent different concepts, growing length scales in the system should follow the same scaling. 
The effective exponents $\beta$ and $\gamma$ for $n(t)$ and $r(t)$ in all cases are related as $\gamma/\beta \approx 0.5$. 
This indeed reflects the scaling $R_{\rm clus} \sim N_{\rm clus}^{0.5}$ found in Figure~\ref{fig:n-r-clusfield}.
In the case of the flexible and semiflexible chains, and in the stars (not shown), no apparent or significant change
is observed in the power-law behavior of $n(t)$ and $r(t)$ at later times when larger clusters are formed, until the former quantities finally saturate when all clusters have merged into a single one and the systems approach the equilibrium fully collapsed state. An overshoot in $n(t)$ and $r(t)$ before saturation is however found in the microgels. This effect is specially pronounced in the diamond network,
and not surprisingly occurs roughly in the same time window as for the normalized domain length $C(t)$ (Figure~\ref{fig:ct-densfield-microgel}d).
Since $r \sim n^{\nu}$ with $\nu \leq 0.5$, the overshoot is much less pronounced in the growing size of the clusters than in the mass aggregation.

In summary, the analysis of clusters reveals that there are no significant differences in the conformations
of the dense regions that are formed during the coarsening of the different macromolecular architectures, but on the rate at which they are formed, which is essentially the same in flexible systems but can be accelerated through bending stiffness in semiflexible systems.

\subsection{Dynamic correlations}

\begin{figure}[t!]
\centering
\includegraphics[width=1\textwidth]{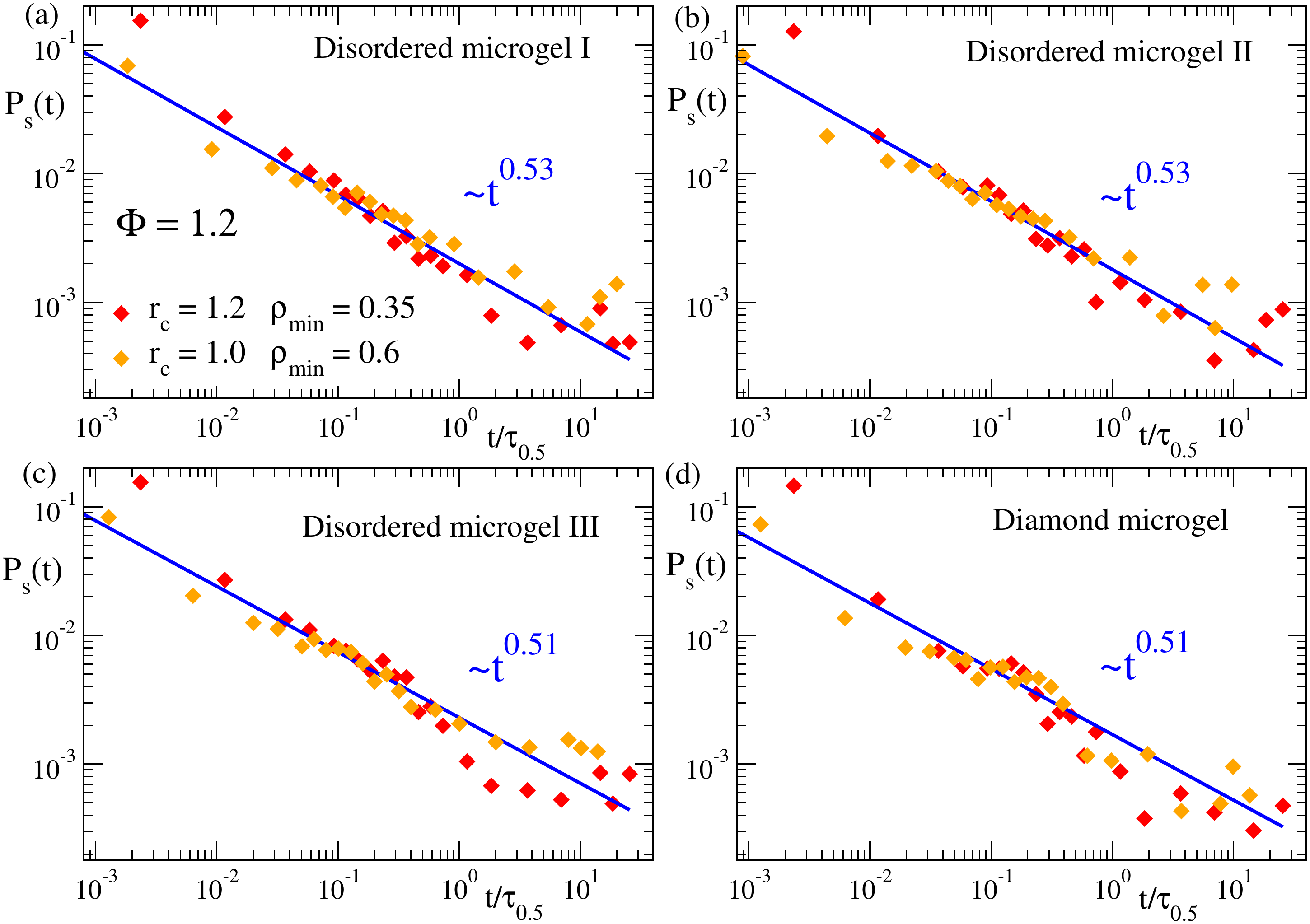}
\caption{Domain self-correlation function $P_{\rm s}(t)$ for three disordered microgels of small (a),
middle (b) and large asphericity (c), and for the diamond-like microgel (d), during the collapse at $\phi = 1.2$. The results displayed correspond to the couples of parameters ($r_{c} = 1.0$, $\rho_{\rm min} = 0.6$) and ($r_{c} = 1.2$, $\rho_{\rm min} = 0.35$)  defining the density field. In each data set the time has been rescaled by $\tau_{0.5}$ , with $\tau_{0.5}$
defined as in Figure \ref{fig:ct-densfield-microgel}. Lines are fits to power-law dependence. Exponents are indicated.}	
\label{fig:correl-microgels} 
\end{figure}

A well-known feature of phase separating and coarsening systems is the scaling of the dynamic correlations with the growing length scale. To test this possibility
we first define a `spin' self-correlation function $P_{\rm s}(t)$ for the density-field representation. The function is defined as \cite{Christiansen2017}:
\begin{equation}
P_{\rm s}(t) = \langle S(t)S(0) \rangle - \langle S(t)\rangle \langle S(0) \rangle
\end{equation}
where the variable $S$ is computed for each cell of the density field, and it is assigned a value 0 or 1 if the cell is empty ($\rho \leq \rho_{\rm min}$) 
or filled ($\rho > \rho_{\rm min}$), respectively.

\begin{figure}[t!]
\centering
\includegraphics[width=.49\textwidth]{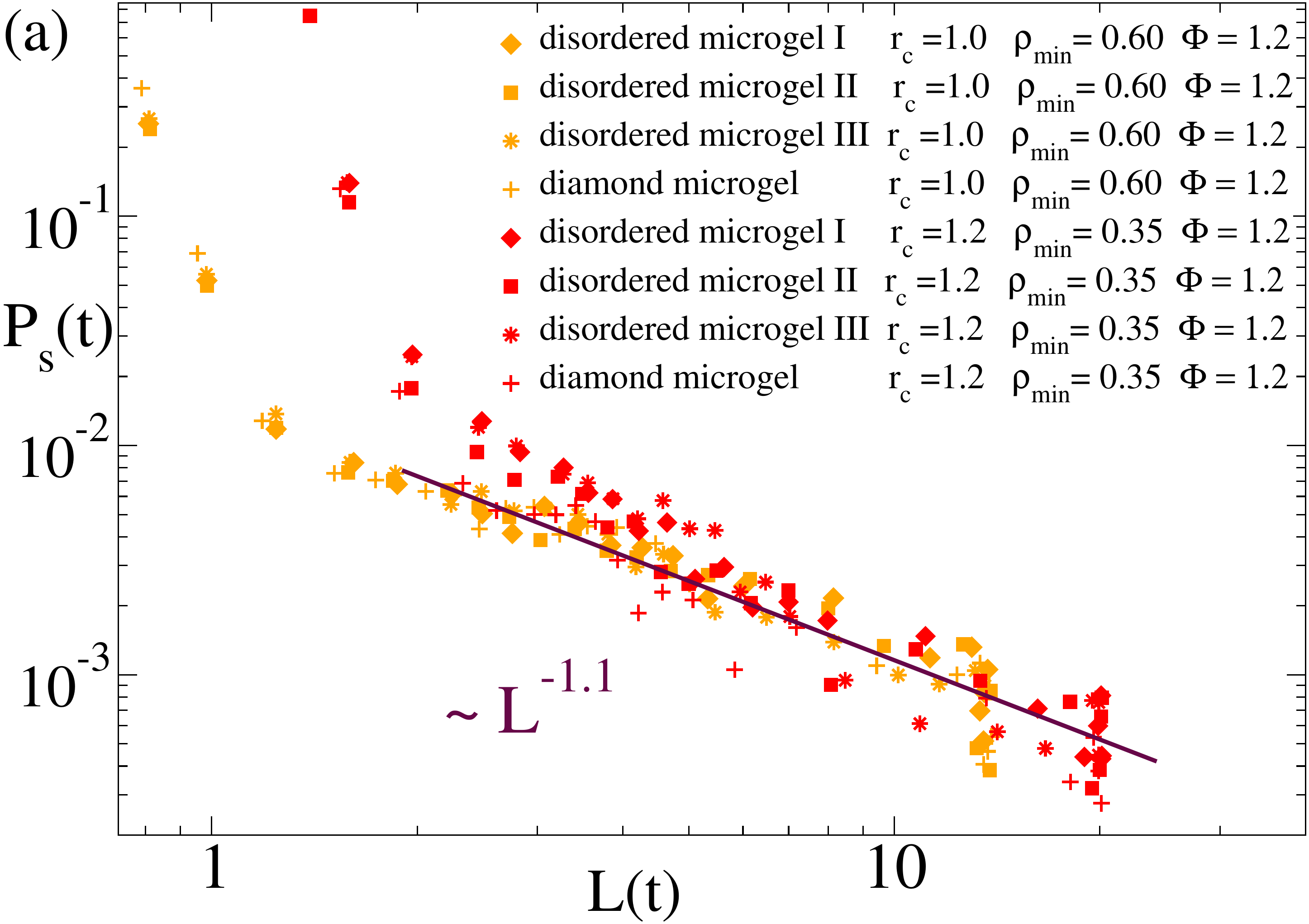}\hspace{0.1cm}
\includegraphics[width=.49\textwidth]{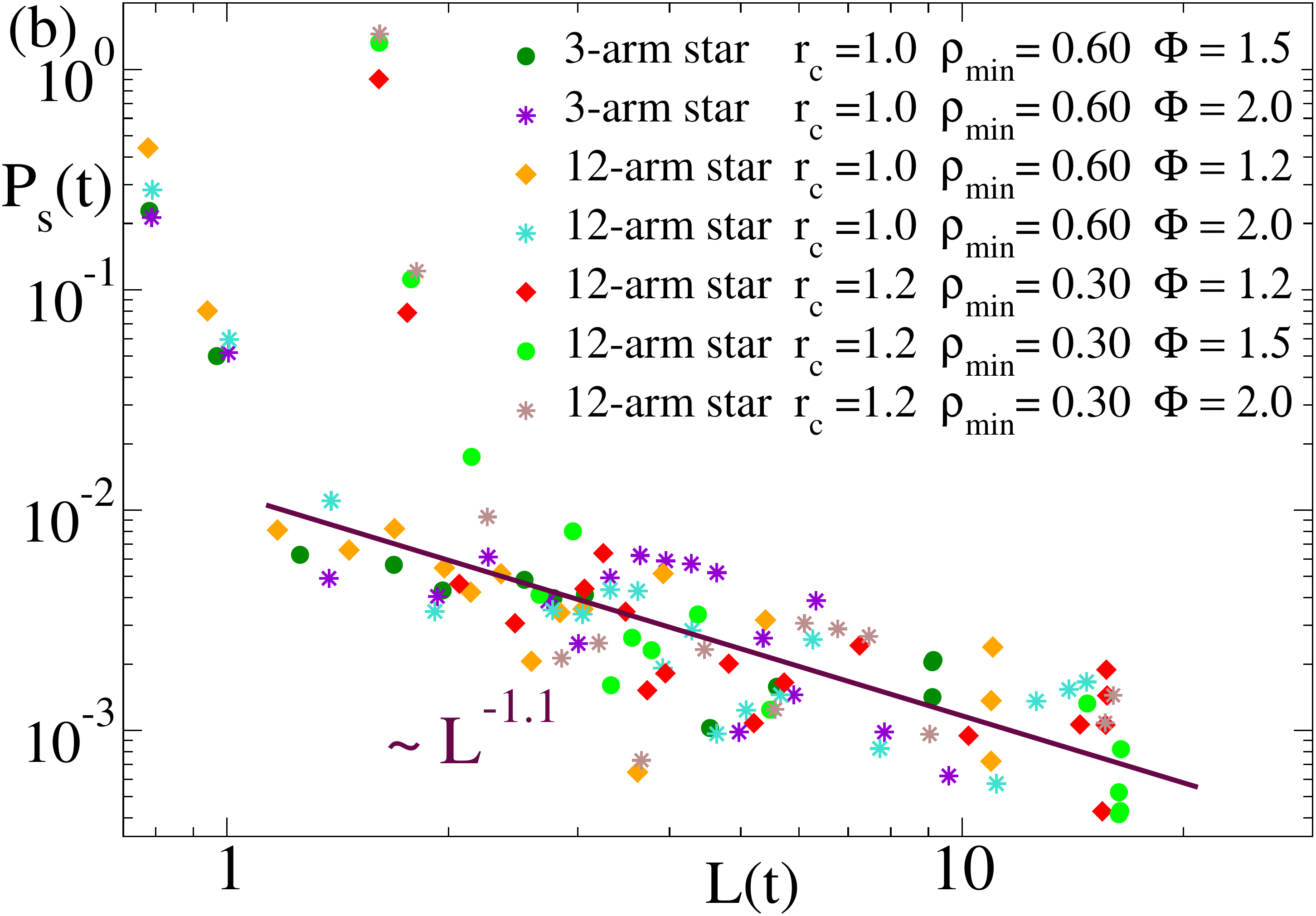}
\caption{Domain self-correlation function $P_{\rm s}(t)$ vs. domain size $L(t)$ at the time $t$  during the
collapse, for several values of the solvent quality parameter $\phi$ and the parameters ($r_{\rm c}$, $\rho_{\rm min}$) defining the density field
(see legends). Panel (a) shows data for the diamond microgel and three disordered microgels.  
Panel (b) shows data for the 3-arm and 12-arm stars.
The lines in both panels are fits to a power-law $P_{\rm s} \sim L^{-1.1}$.}
\label{fig:PvsL} 
\end{figure}

In Figure \ref{fig:correl-microgels} we show the correlation function $P_{\rm s}(t)$ for the three selected disordered microgels (low, middle and high asphericity) 
and for the diamond-like microgel, at solvent parameter $\Phi = 1.2$.
Data are shown for two different couples of the parameters ($r_{\rm c}$, $\rho_{\rm min}$) used to define the density field, and the times are rescaled by $\tau_{0.5}$ as defined above.
Again, the good overlap of the different data sets confirms the consistency of the density field representation for characterizing the coarsening.
The underlying complex dynamics associated to the coarsening process is reflected by the non-exponential decay of the dynamic correlations, which indeed follows a power-law,
$P_{\rm s}(t) \sim t^{-y}$. Results for all the microgels are consistent with an exponent $y \gtrsim 0.5$, irrespective of the network topology.
This is analogous to our results for the domain growth ($C(t)$, Figure~\ref{fig:ct-densfield-microgel}) in the microgels, for which the specific network topology
has no significant effect on the observed power-law.

Moreover, from the computation of $P_{\rm s}(t)$ and the average domain size $L(t)$, we can establish a direct relation between the dynamic self-correlations
and the growing characteristic length scale, just by taking at each time $t$ the corresponding values of $P_{\rm s}$ and $L$.
Figure~\ref{fig:PvsL} shows the results for this relation in the diamond-like microgel and in the three selected disordered microgels (panel (a))
and in the 3-arm and 12-arm stars (panel (b)). 
The data are computed for several values of the solvent parameter $\phi$ and the parameters ($r_{\rm c}$, $\rho_{\rm min}$) used to define the density field.
Time-temperature superposition is also confirmed for this scaling relation, and
data for all flexible systems are consistent with a common power law $P_{\rm s} \sim L^{-1.1}$.
Unfortunately we could not confirm this observation for the semiflexible case, for which the power-law regime in $P_{\rm s}(L)$ developed 
over a short range --- note that the size of the investigated semiflexible chains is much smaller than for the flexible systems --- and had very poor statistics. Confirmation would require to simulate much longer semiflexible chains, which would highly complicate the analysis. 
Tests for much longer chains showed that they tend to collapse into rods and in an
extremely heterogeneous fashion (visiting several long-living intermediate conformations prior to the equilibrium state).

\section{Conclusions}\label{sec5}

By means of simulations, we have investigated the coarsening kinetics emerging during the collapse of several macromolecular architectures in bad solvent:
microgels with realistic (disordered) and with ideal regular (diamond) networks, star polymers of 3 and 12 arms, and linear chains. 
We have also investigated the effect
of bending stiffness on the coarsening kinetics by simulating the collapse of semiflexible linear chains.
In order to remove fast fluctuations that can lead to artifacts in the characterization of the growing length scale, we have made use of a smooth
density field representation of the macromolecules. The domain growth during the coarsening follows a power-law behaviour that is independent of any
reasonable selection of the parameters used to construct the density field (grid size and threshold for defining dense regions). 
The scaling behaviour is independent of the solvent quality parameter, in analogy to time-temperature superposition. 
All flexible systems show approximately the same exponent for the time dependence of the coarsening length scale ($\sim t^{0.6}$). An overshoot is found in the diamond networks
in the late stage of the coarsening, which can be tentatively assigned to the regular distribution of the nodes acting as preferential
nucleating centers and their roughly simultaneous merging when the network is close to the collapsed globular state.
The length scale of coarsening shows a clearly steeper growth in the semiflexible chains ($\sim t^{0.8}$). 
To elucidate the origin of this difference, we have analyzed the clusters of dense regions
formed during the coarsening in the density field representation. The clusters in all systems (flexible or semiflexible) show the same fractal behaviour, i.e.,
their size scales with the mass following the same power-laws. This suggests that
the faster growing length scale in the semiflexible chains just originates from a faster mass diffusion along the chain contour,
and not from a distinct structural feature of the aggregates formed during the coarsening process. 
We have analyzed dynamic correlations, and in analogy with critical phenomena, investigated their dependence on the growing length scale $L$.
We find an apparent common power-law dependence of the correlations ($\sim L^{-1.1}$) for all the flexible systems. 

This work reports, to the best of our knowledge, the first comparison of the coarsening kinetics in bad solvent 
for a broad range of macromolecular architectures, and explores the role of chain stiffness.
As such, it provides a general physical scenario, and a valuable set of results for future theoretical developments
in this, still scarcely studied, fundamental problem with potential applications to e.g., protein folding.

\section*{Acknowledgements}
This research was funded by projects
PGC2018-094548-B-I00 (MCIU/AEI/FEDER, UE) and IT-1175-19 (GV, Spain).


\end{document}